\documentclass[runningheads,table,oribibl]{llncs}
\usepackage{tabularx}
\usepackage{bbm}
\usepackage{dsfont}
\usepackage{appendix}
\usepackage[numbers,sort&compress]{natbib}
\usepackage{footnote}
\usepackage{url}
\usepackage{xcolor}
\usepackage{graphicx}
\usepackage{enumitem}
\usepackage{amsmath}
\usepackage{rotating}
\usepackage{multirow}

\newcolumntype{Y}{>{\centering\arraybackslash}X}

\title{Do you MIND? Reflections on the MIND dataset for research on diversity in news recommendations}
\titlerunning{Do you MIND? Reflections on the MIND dataset}
\author{Sanne Vrijenhoek}
\institute{Institute for Information Law, University of Amsterdam \\ 
\email{s.vrijenhoek@uva.nl}}

\begin{document}

\maketitle

\begin{abstract}
The MIND dataset is at the moment of writing the most extensive dataset available for the research and development of news recommender systems. This work analyzes the suitability of the dataset for research on diverse news recommendations. On the one hand we analyze the effect the different steps in the recommendation pipeline have on the distribution of article categories, and on the other hand we check whether the supplied data would be sufficient for more sophisticated diversity analysis. We conclude that while MIND is a great step forward, there is still a lot of room for improvement. 
\end{abstract}


\section{Introduction}
Engagement with the news has dropped drastically over the last years, with interest dropping from 63\% in 2017 to 51\% in 2022, and a significant number of people avoiding news altogether~\cite{newman2022reuters}. News recommender systems may have a role to play in alleviating these issues by correctly identifying a reader's interest and bringing the right content to the right people. An often-heard criticism on news recommender systems is that they increase the risk of locking users in so-called `filter bubbles', where users are consistently presented with items similar to their preferences or items they have interacted with before. The presence of these filter bubbles and their effects has been hard to prove or disprove, exacerbated by the lack of an exact definition~\cite{michiels2022filter}. The in 2020 published MIND dataset \cite{wu2020mind} is at the moment of writing the largest open source dataset for training and evaluating news recommender systems. It also comes with a set of state-of-the-art news recommender systems that can be trained to predict the articles users will click, and can as such be used to investigate the influence news recommender systems have on the distribution of news content and its diversity, often quoted as the antidote for filter bubbles.
Recent work has argued for a normative interpretation of diversity that reflects the role news plays in democratic society~\cite{vrijenhoek2021recommenders,vrijenhoek2022radio}. The normative diversity metrics proposed here rely on complex metadata that is not readily available without sophisticated analysis of article content. Furthermore, they are mostly tailored towards so-called `hard'  news. In general, \emph{”[F]oreign and domestic politics, economy and finance are usually regarded as hard news. News about sports, celebrities, royal families, crime, scandals and service are regarded as soft news.”}\cite{reinemann2012hard}. In this regard the MIND dataset comes with a number of caveats. MSN News (rebranded Microsoft Start in September 2021) is a news aggregator, and there is very little information available on what news content makes it onto the platform, and how values such as diversity and inclusivity are balanced with financial gains\footnote{according to its Support page \emph{``[...] the content we show aligns with our values and [...] crucial information features prominently in our experiences"})}. As the MIND dataset is expected to contain a significant amount of soft news, it may not be directly useful for research into normative diversity, and experiments run on it may come back skewed. To investigate this we study the overall content present in MIND, using the article category, which is directly available in the dataset, as the relevant unit of analysis. By comparing the presence of article categories at different stages of the recommendation pipeline we can analyze both the influence the recommender system has on the distribution of content and the datasets' suitability for more in-depth news diversity research.

\section{Method}

MIND contains the interactions of 1 million randomly sampled and anonymized users with the news items on MSN News between October 12 and November 22 of 2019. Each datapoint contains an anonymized user id, the user's reading history at that point in time, a list of which items were presented to the user (which we here refer to as the `candidate list'), and which of these items the user ended up clicking. \citet{wu2020mind} describe the performance of several news recommender algorithms when trained on this dataset, including news-specific recommendation methods NPA, NAML, LSTUR and NRMS. The recommenders rank each candidate based on the likeliness a user will click on it. Unfortunately, how the items for the candidate list are chosen is not discussed in the paper. The data is split among training-, validation- and test sets.
We generate the recommendations by running the code in the supplied notebooks\footnote{\url{https://github.com/microsoft/recommenders/tree/main/examples/00_quick_start}} with the large validation set. In a future iteration of this paper, the analysis will be run on the large test set. In total, 376.471 interactions with the system are recorded here, and on average each candidate list consisted of 37 items. 25\% of all interactions had 10 items or less in the candidate list. Close to half (179.383) of the anonymized user ids are unique, with roughly 50.000 and 16.000 ids occurring respectively 2 and 3 times, and 10.000 ids returning more frequently. We assume that user ids are static, and that this means that half of the users only access the site once, and roughly 48.000 visits are from recurring users ($>$4 times). The average time difference between a user id's first- and last recording in the system is 6 hours and 22 minutes, and the maximum 23 hours and 20 minutes. This correlates with an important caveat of the validation set: it only contains data from November 15, 2019.
We calculate the overlap between the generated recommendations using Rank-Biased Overlap (RBO)~\cite{webber2010similarity}, reported in Table \ref{tab:rbo}. This shows a strong overlap in the results of the neural recommenders of $0.61 - 0.63$ between most recommenders, with a much higher score of $0.746$ between NRMS and NAML. As Rank Biased Overlap weighs matches at the beginning of the lists more heavily than those at the end, the recommendations should be divergent enough to observe differences in their outcomes. Interestingly, LSTUR and NRMS are reported by \citep{wu2020mind} to perform best in terms of accuracy, and show comparatively little overlap in Table \ref{tab:rbo}. To avoid redundancy we will only include LSTUR and NRMS in further analysis and comparison with the content in the dataset.

\begin{table}[t]
\caption{Rank-biased Overlap (RBO) between different neural recommender strategies. The calculation encompasses the complete ranking list.}
\centering
\label{tab:rbo}
\begin{tabularx}{\linewidth}{
>{\columncolor[HTML]{FFFFFF}}Y|Y|Y|Y}
               & \textbf{LSTUR} & \textbf{NPA}         & \textbf{NAML}           \\ \hline
\textbf{NRMS} & 0,614	& 0,626 &	0,746 \\
\textbf{NAML}   & 0,616 & 0,639 &	- \\
\textbf{NPA}  & 0,635 &  - & -  \\      
\end{tabularx}
\end{table}

\section{Results}
The different article categories present at different stages of the recommendation pipeline are counted and averaged, the results of which are displayed in Table \ref{tab:table_categories}. For the recommended items, only the top 8 are considered.\footnote{The dataset also contains a few items with categories `kids', `middleeast' and `games', but as these appear less than $0.1\%$ in the full dataset and never in the recommendations they are left out of the analysis.}
As the goal of the neural recommenders is to predict which items have been clicked, the category distributions for the neural recommenders often resembles the distribution in the `clicked' column.

\begin{table}[]
\footnotesize
\caption{Distribution of the different article categories (the whole dataset, what was in the users' reading history, the dataset after candidate selection, and what the user clicked), and the recommender approaches. For the recommendations the top 8 items are selected. The distribution shown does not account for ranking.}
\label{tab:table_categories}
\begin{tabularx}{\linewidth}{l|YYYY|YY}
\textbf{}     & \multicolumn{4}{c|}{\cellcolor[HTML]{FFFFFF}\textbf{MIND}}                                                                    & \multicolumn{2}{c}{\cellcolor[HTML]{FFFFFF}\textbf{Recommendations}} \\

\cline{2-7} 
\rowcolor[HTML]{FFFFFF} 
\multicolumn{1}{l|}{\cellcolor[HTML]{FFFFFF}} &
  all &
  candidate &
  history &
  \multicolumn{1}{c|}{\cellcolor[HTML]{FFFFFF}clicked} &
  LSTUR &
  NRMS \\ \hline
\rowcolor[HTML]{E7E6E6} 
\multicolumn{1}{l|}{\cellcolor[HTML]{E7E6E6}hard} &
  0,363 &
  0,302 &
  0,348 &
  \multicolumn{1}{c|}{\cellcolor[HTML]{E7E6E6}0,269} &
  0,261 &
  0,253 \\
\rowcolor[HTML]{E7E6E6} 
\multicolumn{1}{l|}{\cellcolor[HTML]{E7E6E6}soft} &
  0,636 &
  0,698 &
  0,622 &
  \multicolumn{1}{c|}{\cellcolor[HTML]{E7E6E6}0,730} &
  0,739 &
  0,747 \\
\multicolumn{1}{l|}{\cellcolor[HTML]{FFFFFF}news} &
  \cellcolor[HTML]{F96E6C}0,305 &
  \cellcolor[HTML]{FB9073}0,233 &
  \cellcolor[HTML]{F97A6F}0,279 &
  \multicolumn{1}{c|}{\cellcolor[HTML]{FB8F73}0,235} &
  \cellcolor[HTML]{FB9875}0,215 &
  \cellcolor[HTML]{FB9474}0,224 \\
\multicolumn{1}{l|}{\cellcolor[HTML]{FFFFFF}sports} &
  \cellcolor[HTML]{F8696B}0,314 &
  \cellcolor[HTML]{FCB179}0,163 &
  \cellcolor[HTML]{FDBB7B}0,142 &
  \multicolumn{1}{c|}{\cellcolor[HTML]{FA8A72}0,245} &
  \cellcolor[HTML]{FB9B75}0,209 &
  \cellcolor[HTML]{FB9C75}0,207 \\
\multicolumn{1}{l|}{\cellcolor[HTML]{FFFFFF}finance} &
  \cellcolor[HTML]{FFE383}0,058 &
  \cellcolor[HTML]{FFDD82}0,069 &
  \cellcolor[HTML]{FFDD82}0,069 &
  \multicolumn{1}{c|}{\cellcolor[HTML]{E5E382}0,034} &
  \cellcolor[HTML]{FFE884}0,046 &
  \cellcolor[HTML]{CADB80}0,029 \\
\multicolumn{1}{l|}{\cellcolor[HTML]{FFFFFF}travel} &
  \cellcolor[HTML]{FFE784}0,049 &
  \cellcolor[HTML]{C1D980}0,027 &
  \cellcolor[HTML]{CFDD81}0,030 &
  \multicolumn{1}{c|}{\cellcolor[HTML]{9FCF7E}0,020} &
  \cellcolor[HTML]{B1D47F}0,024 &
  \cellcolor[HTML]{A4D17E}0,021 \\
\multicolumn{1}{l|}{\cellcolor[HTML]{FFFFFF}video} &
  \cellcolor[HTML]{FFE984}0,045 &
  \cellcolor[HTML]{A0CF7E}0,020 &
  \cellcolor[HTML]{98CD7E}0,019 &
  \multicolumn{1}{c|}{\cellcolor[HTML]{A5D17E}0,021} &
  \cellcolor[HTML]{A2D07E}0,021 &
  \cellcolor[HTML]{8AC97D}0,016 \\
\multicolumn{1}{l|}{\cellcolor[HTML]{FFFFFF}lifestyle} &
  \cellcolor[HTML]{FFE984}0,044 &
  \cellcolor[HTML]{FCAD78}0,171 &
  \cellcolor[HTML]{FECC7E}0,105 &
  \multicolumn{1}{c|}{\cellcolor[HTML]{FCAA78}0,178} &
  \cellcolor[HTML]{FCAC78}0,174 &
  \cellcolor[HTML]{FCA777}0,185 \\
\multicolumn{1}{l|}{\cellcolor[HTML]{FFFFFF}foodanddrink} &
  \cellcolor[HTML]{FFE984}0,043 &
  \cellcolor[HTML]{FFDE82}0,068 &
  \cellcolor[HTML]{FFE683}0,050 &
  \multicolumn{1}{c|}{\cellcolor[HTML]{FFE383}0,057} &
  \cellcolor[HTML]{FFD981}0,078 &
  \cellcolor[HTML]{FFDA81}0,077 \\
\multicolumn{1}{l|}{\cellcolor[HTML]{FFFFFF}weather} &
  \cellcolor[HTML]{FFEB84}0,040 &
  \cellcolor[HTML]{C5DA80}0,028 &
  \cellcolor[HTML]{75C37C}0,012 &
  \multicolumn{1}{c|}{\cellcolor[HTML]{85C77C}0,015} &
  \cellcolor[HTML]{C6DA80}0,028 &
  \cellcolor[HTML]{A6D17E}0,021 \\
\multicolumn{1}{l|}{\cellcolor[HTML]{FFFFFF}autos} &
  \cellcolor[HTML]{D0DD81}0,030 &
  \cellcolor[HTML]{CFDD81}0,030 &
  \cellcolor[HTML]{EDE582}0,036 &
  \multicolumn{1}{c|}{\cellcolor[HTML]{B3D57F}0,024} &
  \cellcolor[HTML]{9CCE7E}0,019 &
  \cellcolor[HTML]{9BCE7E}0,019 \\
\multicolumn{1}{l|}{\cellcolor[HTML]{FFFFFF}health} &
  \cellcolor[HTML]{C9DB80}0,028 &
  \cellcolor[HTML]{E5E382}0,034 &
  \cellcolor[HTML]{FFE884}0,047 &
  \multicolumn{1}{c|}{\cellcolor[HTML]{E4E382}0,034} &
  \cellcolor[HTML]{FFE884}0,047 &
  \cellcolor[HTML]{FFEB84}0,041 \\
\multicolumn{1}{l|}{\cellcolor[HTML]{FFFFFF}music} &
  \cellcolor[HTML]{7BC57C}0,013 &
  \cellcolor[HTML]{FFE984}0,044 &
  \cellcolor[HTML]{C2D980}0,027 &
  \multicolumn{1}{c|}{\cellcolor[HTML]{E7E482}0,035} &
  \cellcolor[HTML]{FFEA84}0,042 &
  \cellcolor[HTML]{FFE383}0,057 \\
\multicolumn{1}{l|}{\cellcolor[HTML]{FFFFFF}tv} &
  \cellcolor[HTML]{7BC57C}0,013 &
  \cellcolor[HTML]{FFE884}0,047 &
  \cellcolor[HTML]{FED781}0,082 &
  \multicolumn{1}{c|}{\cellcolor[HTML]{FFE784}0,048} &
  \cellcolor[HTML]{FFE884}0,046 &
  \cellcolor[HTML]{FFEB84}0,041 \\
entertainment &
  \cellcolor[HTML]{63BE7B}0,008 &
  \cellcolor[HTML]{CBDC81}0,029 &
  \cellcolor[HTML]{E7E482}0,034 &
  \cellcolor[HTML]{B0D47F}0,024 &
  \cellcolor[HTML]{CADB80}0,029 &
  \cellcolor[HTML]{E3E382}0,034 \\
movies &
  \cellcolor[HTML]{63BE7B}0,008 &
  \cellcolor[HTML]{F8E983}0,038 &
  \cellcolor[HTML]{FAE983}0,038 &
  \cellcolor[HTML]{CEDD81}0,030 &
  \cellcolor[HTML]{B0D47F}0,024 &
  \cellcolor[HTML]{C6DA80}0,028
\end{tabularx}
\end{table}

\label{sec:distribution}
Table \ref{tab:table_categories} shows a general overview of the distribution of article categories among all the articles that were in the dataset, the result after the first candidate selection, what was in users' history, and in the set of articles recommended by LSTUR and NRMS. Furthermore, we aggregate the categories present into \emph{hard} and \emph{soft} news following the distinction described in the Introduction. In this dataset, this means that the categories `news' and `finance' are considered hard news, whereas the rest is soft. 
One major discrepancy can already be observed after candidate selection: the `lifestyle' category, which in the complete dataset only accounts for 4.4\% of the articles, has a comparatively big representation (17\%) in the set of candidates. The news and sport categories are the most inversely affected, with a 30\% and 31\% representation in the overall dataset and 23\% and 16\% after candidate selection. Because the recommender strategies evaluated here have no influence over the candidate selection, this is an important observation to take into account. 

In general, the two news-specific recommender strategies seem to behave largely similar. Given that the neural recommenders take the items in users' reading history into account, we would expect the recommenders to reflect similar patterns as the history; however, this does not seem to be the case. On the contrary, while the list of candidate items consisted of 23\% news items, and the reading history almost 28\%, the neural recommenders are further downplaying the share of news items in the recommendations, containing only about 22\% news. The opposite happens for the sport and lifestyle category: where the candidate selection contains 16\% sport and 17\% lifestyle, and the reading history respectively 14\% and 10\%, the LSTUR recommender is increasing the presence of these categories to 21\% and 17\%. It does, however, very closely resemble the distribution of items that users have clicked, which is also not surprising given that this is what the recommender is optimized on.  

More interesting patterns can be observed when considering the length of the recommendation, as shown in Table \ref{tab:n}. At 1, only the item with the highest predicted relevance is included, continuing on until all items in the recommendations are. At this point, the recommendation is equal to the full candidate list, as ordering is not taken into account in this analysis. The table is ordered on the category's share in position 1, which is of extra importance given an average user's tendency towards position bias~\cite{positionbias}. Both LSTUR and NRMS are very likely to recommend sports and news at the beginning of a recommendation. Finance only appears much later: despite it's relatively large presence in both the overall dataset and the candidate list (7\% and 6\%), finance does not even appear among NRMS' top 10 categories. They also both prominently feature category `foodanddrink' in first position (10\% and 15\%, versus only 7\% in the candidate list). But we see also more distinct differences: NRMS comparatively often recommends items from it's top categories in first place, whereas for LSTUR this is more spread out. At the first position, NRMS recommends more than 83\% of the content out of 4 most frequently occurring categories, whereas for LSTUR this is around 74\%. NRMS is also much more likely than LSTUR to recommend news in the first position (28\% vs. 22\%). At position 10, both recommenders actually list \emph{less} news than in the overall dataset. It seems here that news either gets recommended in the top positions, or not at all. 


\begin{table}[]
\caption{Distribution of the top 8 article categories at different recommendation lengths, ordered by frequency at recommendation length 1. Category `food' is short for category `foodanddrink'.}
\label{tab:n}
\begin{tabularx}{\linewidth}{cl|YYYYYY}
\multicolumn{1}{l}{\cellcolor[HTML]{FFFFFF}}             & \multicolumn{1}{l|}{\cellcolor[HTML]{FFFFFF}\textbf{}} & \multicolumn{1}{c}{\cellcolor[HTML]{FFFFFF}\textbf{1}} & \multicolumn{1}{c}{\cellcolor[HTML]{FFFFFF}\textbf{2}} & \multicolumn{1}{c}{\cellcolor[HTML]{FFFFFF}\textbf{5}} & \multicolumn{1}{c}{\cellcolor[HTML]{FFFFFF}\textbf{10}} & \multicolumn{1}{c}{\textbf{20}} & \multicolumn{1}{c}{\textbf{$\infty$}} \\ \hline
\multicolumn{1}{c|}{\cellcolor[HTML]{E7E6E6}}                                 & \multicolumn{1}{l|}{sports}                            & \cellcolor[HTML]{FA8270}0,24       & \cellcolor[HTML]{FA8170}0,25       & \cellcolor[HTML]{FA8C72}0,23       & \cellcolor[HTML]{FB9C75}0,20        & \cellcolor[HTML]{FCA877}0,18 & \cellcolor[HTML]{FCB37A}0,16 \\
\multicolumn{1}{c|}{\cellcolor[HTML]{E7E6E6}}                                 & \multicolumn{1}{l|}{\cellcolor[HTML]{EFEFEF}news}      & \cellcolor[HTML]{FB8F73}0,22       & \cellcolor[HTML]{FB9474}0,21       & \cellcolor[HTML]{FB9474}0,21       & \cellcolor[HTML]{FB9373}0,22        & \cellcolor[HTML]{FB8F73}0,22 & \cellcolor[HTML]{FA8871}0,23 \\
\multicolumn{1}{c|}{\cellcolor[HTML]{E7E6E6}}                                 & \multicolumn{1}{l|}{\cellcolor[HTML]{EFEFEF}lifestyle} & \cellcolor[HTML]{FCAF79}0,17       & \cellcolor[HTML]{FCAC78}0,17       & \cellcolor[HTML]{FCAC78}0,17       & \cellcolor[HTML]{FCAC78}0,17        & \cellcolor[HTML]{FCAD79}0,17 & \cellcolor[HTML]{FCAD79}0,17 \\
\multicolumn{1}{c|}{\cellcolor[HTML]{E7E6E6}}                                 & \multicolumn{1}{l|}{\cellcolor[HTML]{EFEFEF}food}      & \cellcolor[HTML]{FED781}0,10       & \cellcolor[HTML]{FFDC82}0,09       & \cellcolor[HTML]{FFE383}0,08       & \cellcolor[HTML]{FFE784}0,08        & \cellcolor[HTML]{FFE984}0,07 & \cellcolor[HTML]{FFEB84}0,07 \\
\multicolumn{1}{c|}{\cellcolor[HTML]{E7E6E6}}                                 & \multicolumn{1}{l|}{\cellcolor[HTML]{EFEFEF}health}    & \cellcolor[HTML]{AAD27F}0,04       & \cellcolor[HTML]{B4D57F}0,05       & \cellcolor[HTML]{BBD780}0,05       & \cellcolor[HTML]{BAD780}0,05        & \cellcolor[HTML]{B2D47F}0,04 & \cellcolor[HTML]{91CB7D}0,03 \\
\multicolumn{1}{c|}{\cellcolor[HTML]{E7E6E6}}                                 & \multicolumn{1}{l|}{travel}                            & \cellcolor[HTML]{A5D17E}0,04       & \cellcolor[HTML]{8DCA7D}0,03       & \cellcolor[HTML]{76C37C}0,03       & \cellcolor[HTML]{6DC17B}0,02        & \cellcolor[HTML]{6EC17B}0,02 & \cellcolor[HTML]{79C47C}0,03 \\
\multicolumn{1}{c|}{\cellcolor[HTML]{E7E6E6}}                                 & \multicolumn{1}{l|}{enter}                             & \cellcolor[HTML]{8FCA7D}0,03       & \cellcolor[HTML]{8AC97D}0,03       & \cellcolor[HTML]{83C77C}0,03       & \cellcolor[HTML]{7DC57C}0,03        & \cellcolor[HTML]{7BC47C}0,03 & \cellcolor[HTML]{80C67C}0,03 \\
\multicolumn{1}{c|}{\multirow{-8}{*}{\rotatebox[origin=c]{90}{\cellcolor[HTML]{E7E6E6}\textbf{LSTUR}}}} & \multicolumn{1}{l|}{finance}                           & \cellcolor[HTML]{8BC97D}0,03       & \cellcolor[HTML]{94CC7D}0,04       & \cellcolor[HTML]{A6D17E}0,04       & \cellcolor[HTML]{C1D980}0,05        & \cellcolor[HTML]{E1E282}0,06 & \cellcolor[HTML]{FFEB84}0,07 \\ \hline
\multicolumn{1}{c|}{\cellcolor[HTML]{E7E6E6}}                                 & \multicolumn{1}{l|}{news}                              & \cellcolor[HTML]{F8696B}0,28       & \cellcolor[HTML]{F97A6F}0,26       & \cellcolor[HTML]{FA8972}0,23       & \cellcolor[HTML]{FB8F73}0,22        & \cellcolor[HTML]{FB9073}0,22 & \cellcolor[HTML]{FA8871}0,23 \\
\multicolumn{1}{c|}{\cellcolor[HTML]{E7E6E6}}                                 & \multicolumn{1}{l|}{sports}                            & \cellcolor[HTML]{FB9273}0,22       & \cellcolor[HTML]{FB9073}0,22       & \cellcolor[HTML]{FB9173}0,22       & \cellcolor[HTML]{FB9C75}0,20        & \cellcolor[HTML]{FCA777}0,18 & \cellcolor[HTML]{FCB37A}0,16 \\
\multicolumn{1}{c|}{\cellcolor[HTML]{E7E6E6}}                                 & \multicolumn{1}{l|}{lifestyle}                         & \cellcolor[HTML]{FCB179}0,17       & \cellcolor[HTML]{FCAB78}0,18       & \cellcolor[HTML]{FCA677}0,18       & \cellcolor[HTML]{FCA677}0,18        & \cellcolor[HTML]{FCA978}0,18 & \cellcolor[HTML]{FCAD79}0,17 \\
\multicolumn{1}{c|}{\cellcolor[HTML]{E7E6E6}}                                 & \multicolumn{1}{l|}{food}                              & \cellcolor[HTML]{FDB87B}0,15       & \cellcolor[HTML]{FDC67D}0,13       & \cellcolor[HTML]{FFDD82}0,09       & \cellcolor[HTML]{FFE984}0,07        & \cellcolor[HTML]{FFEB84}0,07 & \cellcolor[HTML]{FFEB84}0,07 \\
\multicolumn{1}{c|}{\cellcolor[HTML]{E7E6E6}}                                 & \multicolumn{1}{l|}{music}                             & \cellcolor[HTML]{B4D57F}0,04       & \cellcolor[HTML]{D8E081}0,06       & \cellcolor[HTML]{E4E382}0,06       & \cellcolor[HTML]{D9E081}0,06        & \cellcolor[HTML]{C6DA80}0,05 & \cellcolor[HTML]{B2D47F}0,04 \\
\multicolumn{1}{c|}{\cellcolor[HTML]{E7E6E6}}                                 & \multicolumn{1}{l|}{enter}                             & \cellcolor[HTML]{7BC47C}0,03       & \cellcolor[HTML]{83C77C}0,03       & \cellcolor[HTML]{8DCA7D}0,03       & \cellcolor[HTML]{90CB7D}0,03        & \cellcolor[HTML]{8CC97D}0,03 & \cellcolor[HTML]{80C67C}0,03 \\
\multicolumn{1}{c|}{\cellcolor[HTML]{E7E6E6}}                                 & health                                                 & \cellcolor[HTML]{6EC17B}0,02       & \cellcolor[HTML]{7DC57C}0,03       & \cellcolor[HTML]{9BCE7E}0,04       & \cellcolor[HTML]{ACD37F}0,04        & \cellcolor[HTML]{AED37F}0,04 & \cellcolor[HTML]{91CB7D}0,03 \\
\multicolumn{1}{c|}{\multirow{-8}{*}{\rotatebox[origin=c]{90}{\cellcolor[HTML]{E7E6E6}\textbf{NRMS}}}}  & tv                                                     & \cellcolor[HTML]{63BE7B}0,02       & \cellcolor[HTML]{70C17B}0,02       & \cellcolor[HTML]{90CB7D}0,03       & \cellcolor[HTML]{B1D47F}0,04        & \cellcolor[HTML]{C1D980}0,05 & \cellcolor[HTML]{BBD780}0,05
\end{tabularx}
\end{table}

\section{Conclusion}
Analyzing the results of the different recommendation strategies reveals characteristics of the recommendations that are not visible when purely reporting on performance statistics such as NDCG or AUC. The neural recommenders have a distinct impact on the dissemination of content, especially considering what content is present in the overall dataset and the type of content users have clicked in the past. As expected, the neural recommenders largely reflect the types of clicks that have been recorded. The candidate list reduces the presence of frequently occurring categories while inflating that of less frequent ones. However, this behavior is to be expected; the candidate selection ought to contain a wide range of content, so that the recommender system can correctly identify content that is specifically relevant to that particular user. It does however raise questions about the granularity of the categories chosen. In the design of MIND the choice was made to distinguish between `movies', `music', `tv' and `entertainment', even though these account for less than 10\% of all items in the dataset. A dataset that is more focused on news content could instead split this top-level category into subcategories such as `local news', `global news' or `politics'. MIND does contain subcategories, such as `newsus', `newspolitics', and `newsworld' (respectively 47\%, 17\% and 8\% of all news items), which could be more relevant for future research.  

The neural recommenders also behave differently when compared to each other, with NRMS prominently recommending news and food in top positions, and LSTUR favoring sports and other, less common categories.  With the neural recommenders largely focusing on lifestyle and entertainment, and downplaying news and finance, one could argue they mostly promote soft news. This is not to say these personalized recommendations are bad; there can be value in bringing the right soft news to the right people, as ~\citet{andersen2019entrance} notes that consuming soft news may serve as a stepping stone to more active participation and engagement. This does warrant a more in-depth discussion about the purpose of the recommender system, a thorough investigation into the mismatch between produced content, user reading history and user clicking behavior, and an editorial decision on the balance between `quality' and `fun'~\cite{smets2022we}. 

In terms of research into normative diversity, MIND leaves a few things to be desired. With only 20\% of articles in the recommendations being news articles, there is only little information to determine whether users receive a balanced overview of the news. This is strengthened by the lack of metadata that is present in the dataset: only the article title, (sub)category and url are directly supplied. Automatic stance- or viewpoint detection based on article fulltext, which could be retrieved by following the url to the MSN News website, may be a direction for future research~\cite{reuver-etal-2021-human}. For example, \citet{mascarell2021stance} published a detailed annotation of different stances and emotions present in German news articles. They do, however, lack the scale and user interactions that MIND has. 

The majority of interactions recorded in MIND are (assumed to be) unique visits, though it does contain a considerable amount of returning users: almost 10.000 access the system 4 times or more, resulting in a total of 48.000 visits from recurring users. If we combine this with the average length of the candidate list of 37 items, and the fact that 22\% of recommended items is news, this yields us about 400.000 news items shown. However, even when the users return to the system more frequently, the validation set only contains information on the interactions users had with the system on one specific day, making it impossible to see how the users and the recommender's behavior towards those users change over time \cite{michiels2022filter}. While the large test set does contain data over six days rather than just one, this would still not be enough to actually see differences in users' behavior, even if they do use the system intensively.  
Ideally, if one were to research the effect of a recommender system on the diversity of consumed news, they would want to do this based on a system with 1) a large number of frequently returning users (though a smaller number of unique users compared to MIND would be acceptable), 2) a focus on hard news, and 3) over a longer period of time, allowing for both the users and the recommender system to evolve over time.  
In conclusion: the MIND dataset is, especially given the fact that it is open source, a great step forward in the research on news recommender systems and their effects. However, when to goal is to move the discussion beyond recommender accuracy and towards news recommender diversity, there are still several points of improvement necessary. 

\section*{Acknowledgements}
I thank Mateo Gutierrez Granada for his help in generating the recommendations used in this analysis. I also thank Savvina Daniil, Lien Michiels and an anonymous reviewer for their critical comments on earlier versions of this work, and in doing so their contributions to improving the end product. 

\bibliographystyle{splncs04}
\bibliography{bibliography}

\end{document}